 \documentclass[final,5p,times,authoryear]{elsarticle}

\usepackage{amssymb}
\usepackage{lipsum}
\usepackage{multirow}

\journal{Astronomy $\&$ Computing}

\begin{document}

\begin{frontmatter}



\title{Processing of GASKAP-H\textsc{i} pilot survey data using a commercial supercomputer}

\author[one,two]{I. P. Kemp\corref{cor1}}
\ead{ian.kemp@postgrad.curtin.edu.au}
\cortext[cor1]{Corresponding author.}
\affiliation[one]{organization={International Centre for Radio Astronomy Research (ICRAR)}, addressline={Curtin University, Bentley, 6102, WA, Australia}}

\affiliation[two]{organization={CSIRO Space and Astronomy}, addressline={26 Dick Perry Avenue, Kensington, 6151, WA, Australia}}

\author[three]{N. M. Pingel}
\affiliation[three]{organization={University of Wisconsin-Madison}, addressline={4554 Sterling Hall, 475 N. Charter Street, Madison, 53706-1507, WI, USA}}

\author[four]{R. Worth}
\affiliation[four]{organization={DUG Technology}, addressline={76 Kings Park Road, West Perth, 6005, WA, Australia}}

\author[four]{J. Wake}
            
\author[five]{D. A. Mitchell}
\affiliation[five]{organization={CSIRO Space and Astronomy},addressline={Cnr Vimiera \& Pembroke Roads, Marsfield, 2122, NSW, Australia}}

\author[six]{S. D. Midgely}
\affiliation[six]{organization={Defence Science and Technology Group}, addressline={Australian Department of Defence, Australia}}

\author[one]{S. J. Tingay}

\author[seven,eight]{J. Dempsey}
\affiliation[seven]{organization={Research School of Astronomy and Astrophysics, The Australian National University}, addressline={Canberra ACT 2611, Australia}}

\affiliation[eight]{organization={CSIRO Information Management and Technology}, addressline={GPO Box 1700 Canberra, ACT 2601, Australia}}

\author[nine]{H. Dénes}
\affiliation[nine]{organization={School of Physical Sciences and Nanotechnology, Yachay Tech University}, addressline={Hacienda San José S/N, 100119, Urcuquí, Ecuador}}

\author[ten]{J. M. Dickey}
\affiliation[ten]{organization={School of Natural Sciences, University of Tasmania}, addressline={TAS 7005, Australia}}

\author[eleven]{S. J. Gibson}
\affiliation[eleven]{organization={Physics and Astronomy, Western Kentucky University}, addressline={Bowling Green, KY, USA}}

\author[twelve]{K. E. Jameson}
\affiliation[twelve]{organization={Caltech}, addressline={1200 E California Blvd, Pasadena, CA 91125, USA}}

\author[seven]{C. Lynn}

\author[seven]{Y. K. Ma}

\author[seven]{A. Marchal}

\author[seven]{N. M. McClure-Griffiths}

\author[three]{S. Stanimirović}

\author[thirteen]{J. Th. van Loon}
\affiliation[thirteen]{organization={Lennard-Jones Laboratories, Keele University}, addressline={ST5 5BG, UK}}

\begin{abstract}
Modern radio telescopes generate large amounts of data, with the next generation Very Large Array (ngVLA) and the Square Kilometre Array (SKA) expected to feed up to 292 GB of visibilities per second to the science data processor (SDP). However, the continued exponential growth in the power of the world’s largest supercomputers suggests that for the foreseeable future there will be sufficient capacity available to provide for astronomers' needs in processing ‘science ready’ products from the new generation of telescopes, with commercial platforms becoming an option for overflow capacity. The purpose of the current work is to trial the use of commercial high performance computing (HPC) for a large scale processing task in astronomy, in this case processing data from the GASKAP-H\textsc{i} pilot surveys. We delineate a four-step process which can be followed by other researchers wishing to port an existing workflow from a public facility to a commercial provider. We used the process to provide reference images for an ongoing upgrade to ASKAPSoft (the ASKAP SDP software), and to provide science images for the GASKAP collaboration, using the joint deconvolution capability of WSClean. We document the approach to optimising the pipeline to minimise cost and elapsed time at the commercial provider, and give a resource estimate for processing future full survey data. Finally we document advantages, disadvantages, and lessons learned from the project, which will aid other researchers aiming to use commercial supercomputing for radio astronomy imaging. We found the key advantage to be immediate access and high availability, and the main disadvantage to be the need for improved HPC knowledge to take best advantage of the facility.
\end{abstract}




\end{frontmatter}




\section{Introduction}
\label{introduction}

Modern radio interferometer arrays generate large amounts of data. The Science Data Processor (SDP) for the 36-dish ASKAP telescope was designed to process 3 GB per second of visibility data \citep{2019ASPC..521..276G}. The ngVLA, the planned next generation VLA, is expected to produce a peak rate of 133 GB per second \citep{ngVLA-Req}, while SKA-Mid, currently under construction \citep{SKA}, will produce 292 GB per second \citep{2022SPIE12182E..0QM}.

Because of these large data volumes, radio astronomy is critically dependent upon high performance computing (HPC) to carry out imaging and other analysis to extract useful science, such as for example the calibrated and mosaiced image cubes for the WALLABY survey \citep{2020Ap&SS.365..118K}, the atomic neutral hydrogen absorption measurements described in \cite{2022PASA...39...34D}, and the de-dispersed real-time FRB search of the CRAFT survey \citep{2024arXiv240802083S}.

In 2011, \citet{Norris2011} outlined the SKA Data Challenge, and estimated that in 2022, SKA data processing would require computing infrastructure beyond the capacity of the largest supercomputer in the world. In reality, Norris’ concerns have been mitigated by delays to the start of SKA construction. Now, 13 years later, computing availability has already fulfilled the needs of the SKA. This was confirmed by \cite{2020SciBu..65..337W} who successfully carried out processing of simulated SKA data in better than real-time, using Summit, then identified in the Top500 list of supercomputing systems \citep{TOP500} as the world’s most performant computer, with a maximal achievable benchmark performance ($R_{max}$) of 149 PFlop/s. In 2024, just five years after that work was carried out, Summit has been decommissioned, and the current top listed system, El Capitan, is rated at $R_{max}$ = 1742 PFlop/s, 11 times the power of Summit. 

Current plans for the SKA are to carry out a number of standardised workflows at the SDP, with this central facility providing a range of science ready products including image cubes and spectra \citep{2014era..conf20201N}.  Specialist or experimental processing by other users of the data will be carried out using `science regional centres' (SRCs). The data post-processing system built for the Australian SRC has been designed to be hardware agnostic \citep{AusSRC-DSP-Final}, so that in the future it could be implemented using dedicated facilities, national computing infrastructure, or commercially provided HPC or cloud facilities.

This approach has been foreshadowed in the processing of data from the Australian SKA Pathfinder (ASKAP) \citep{2012SPIE.8444E..2AS, 2021PASA...38....9H}. Each of the 36 ASKAP antennas are equipped with 188-element phase array feed receivers, enabling a $\sim$25 deg$^2$ field-of-view through beamforming 36 simultaneous and separate primary beams, which each extend one degree on the sky at 21 cm. The philosophy at ASKAP is to produce and provide `science-ready' data products, including calibrated visibilities and images, using specific software named ASKAPSoft \citep{2020ASPC..527..591W}. The SDP is implemented on Setonix, which is currently the highest performance public HPC system in Australia (positioned 28 on the Top500 list), with an $R_{max}$ of 27 PFlop/s, located at the Pawsey Supercomputing Research Centre in Perth\footnote{https://pawsey.org.au/}. The data rate from the observatory ranges up to 2.5 GB/s \citep{ATNF-ASKAP}, depending on whether ASKAP is observing in the 288 MHz wide spectral-band continuum mode or its various zoom-modes for spectral line science. These large data volumes from ASKAP present a unique opportunity to explore how to efficiently utilise the computing resources available to the astronomical community, in preparation for future facilities.

One of the science survey projects approved for the ASKAP telescope is GASKAP-H\textsc{i}, a survey to study the distribution and thermodynamic properties of atomic neutral hydrogen (H\textsc{i}) in the Milky Way and nearby Magellanic System, at high angular and spectral resolution \citep{dickey2013}. Specific science goals of this survey include characterizing interstellar turbulence in this key component of the interstellar medium and revealing the phase transitions between the multi-phase atomic gas and star-forming molecular gas. In preparation for the main survey, shallow (10$-$20 hour) Phase I and Phase II pilot surveys - capturing the Small Magellanic Cloud (SMC), Large Magellanic Cloud (LMC), Magellanic Bridge, and the start of the Magellanic Stream - have been carried out to develop the imaging and analysis tools, and data from the pilots have already been used to provide useful scientific outputs in line with the survey aims (see \cite{2022PASA...39....5P}, \cite{2022ApJ...926..186D}, \cite{2022PASA...39...34D} \cite{2024MNRAS.534.3478N}).

For GASKAP-H\textsc{i}, the novel phased array feed (PAF) receivers on ASKAP expand the instantaneous field-of-view of the telescope from 1$^{\circ}\times1^{\circ}$ expected for a 12 m dish to $\sim5^{\circ}\times5^{\circ}$ through forming 36 simultaneous primary beams on the sky. The standard approach to imaging a single ASKAP PAF footprint is a linear mosaic, wherein distinct deconvolved images, (i.e., with the inherent PSF response from the sky-brightness distribution removed), produced from each individual beam, are combined in a pointed mosaic. However, in the case of GASKAP-H\textsc{i}, diffuse emission from the Milky Way and nearby Magellanic system extend far beyond the boundaries of individual beams. Because common deconvolution algorithms, such as CLEAN \citep{1974A&AS...15..417H}, are non-linear processes, we risk losing information about the largest scales due to variations in the beam-to-beam sky models. To ensure the largest scale emission is accurately recovered across beam boundaries, a joint deconvolution approach is necessary - in which the images from each beam are stitched together in a single image before moving on to deconvolution. However, this imaging approach is computationally intensive, so that Pawsey's Galaxy supercomputer, the predecessor to Setonix, did not possess sufficient node memory, hampering efforts to develop this imaging mode in ASKAPSoft. Fortunately, the increased capabilities of Setonix has enabled development of joint deconvolution in ASKAPSoft. 

Pending the enhancements to ASKAPSoft, \citet{2022PASA...39....5P} developed an imaging pipeline for GASKAP-H\textsc{i} based on the command-line imaging software WSClean \citep{2014MNRAS.444..606O}. Due to the limited availability of systems with suitable memory to process this data, it is worthwhile to identify alternative computing resources. Of interest here is the commercial sector, which has had a steadily increasing presence in the afore-mentioned Top500 list of supercomputing systems. The June 2024 Top500 list showed that five of the top 20 publicly-disclosed systems were owned by commercial providers. Five years ago the corresponding number was zero - in June 2019 only one machine in the top 20 was owned by a private corporation, and this was an in-house system for an oil company (Total) and not available to the public research community. In the future it may be increasingly possible for commercial supercomputing to provide additional on-demand capacity for radio astronomy data processing.

The aim of the present work is to trial the use of commercial supercomputing, by processing data from the GASKAP-H\textsc{i} pilot surveys using a supercomputer owned by DUG Technology in Perth, Western Australia\footnote{https://dug.com/about-dug/}. The work was part of a larger trial facilitated by CSIRO\footnote{https://www.csiro.au/}, in which teams from multiple disciplines were given access to DUG facilities, to evaluate the applicability of commercial technology. We used DUG's commercial offering, which includes a high level of technical support, in assisting with configuration and optimisation of the deployment of our specialist software. 

Our trial had four sub-objectives:
\begin{enumerate}
\label{objectives}
    \item{Identify and document the feasibility, advantages and disadvantages of using a commercial facility for data processing;}
    \item{Provide a resource estimate for processing full survey data using commercial supercomputing;}
    \item{Use the joint deconvolution technique with WSClean to produce reference images for the updated joint deconvolution algorithm in ASKAPSoft; and}
    \item{Produce fully processed cubes from the pilot survey data, for use by the science team.}
\end{enumerate}

We wish to use our experience to inform the community, so that astronomers may be better placed to take advantage of the large quantities of commercial HPC which we expect to enter the market in the next decade.  Our experience may also help inform the discussion about the prospects for incorporation of commercial supercomputing in the model for the SKA regional centres.

We adapted the existing imaging workflow which has been described in \cite{2022PASA...39....5P}, and in this paper we will focus on the issues around porting the workflow and use of commercial computing facilities.

The structure of this paper is as follows: in Section \ref{Section:Observations} we record the observations used; in Sections \ref{Section:hardware} and \ref{Section:software} we describe the computational infrastructure and software respectively; and in Section \ref{Section:Workflow} we describe the overall processing steps. In Section \ref{Section:Methods} we explain how the workflow was evolved to make optimum use of the commercial facility; in Section \ref{Section:Results} we summarise our results; and in Section \ref{Section:Lessons} we detail the lessons learned from the process.

\section{Observations and data}
\label{Section:Observations}
We used calibrated visibilities from the GASKAP-H\textsc{i} Phase I and Phase II pilot surveys performed with the ASKAP telescope. Primary beams obtained using the holography methods outlined by \cite{2021PASA...38....9H} at 1.4 GHz were used as the `beam\_map' input for the joint deconvolution mode in WSClean. Processing resulted in images with 30$''$ synthesised beams with rms noise of $\sim$3 K per 0.24 km s$^{-1}$ channel. Observations were obtained using 3 interleaved positions, to provide uniform sensitivity data across 108 individual beam pointings for each field. Data were collected in the spectral window 1418.501 MHz - 1420.944 MHz, in 2112 channels of width 1.157 kHz.

\begin{figure*}
\scalebox{.64}{\includegraphics{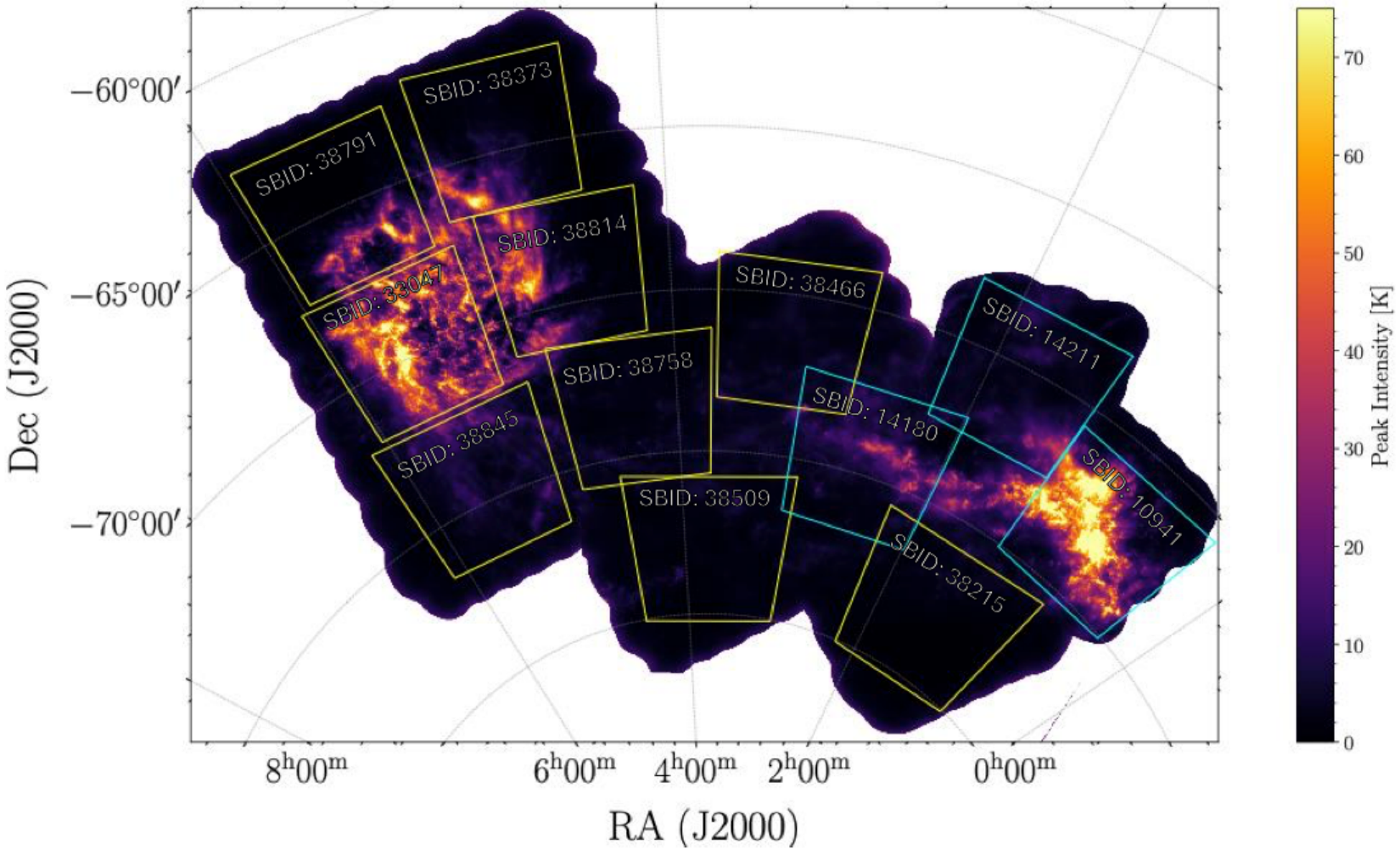}}
\caption{Fields in the GASKAP-H\textsc{i} Pilot Surveys. Cyan boxes denote the approximate ASKAP PAF footprint for the pilot phase I fields (20hr integration); yellow boxes are those for the pilot phase II fields (10hr integration)}
\label{fig:GASKAP_fields}
\end{figure*}

The pilot I survey consisted of 20 hr integrations spread across three fields centered on the Small Magellanic Cloud (SMC), a portion of the Magellanic Bridge, and the beginning of the Magellanic Stream extending from the SMC. The pilot II survey consisted of 10-hour observations across 9 fields across the Magellanic system (see Figure \ref{fig:GASKAP_fields}). In total, each integration generated 108 $\times$ 61 GB, $\sim$6600 GB of visibilities. These visibilities had been calibrated with the standard ASKAPSoft pipeline, as described in Section 3 of \citet{2022PASA...39....5P}. In the work reported here, each field was imaged and deconvolved using the joint deconvolution mode of the WSClean software package \citep{2014MNRAS.444..606O} to produce image cubes of up to 150 GB in size. 

The primary repository for ASKAP data is the CSIRO ASKAP Science Data Archive (CASDA) \citep{2020ASPC..522..263H}. In CASDA, the data are located and referred to using the Scheduling Block IDs (SBID). In this work we used data from SBID 10941 from pilot phase I, and SBIDs 38509, 38466, 38215 from pilot phase II (see Figure \ref{fig:GASKAP_fields}).

\section{Hardware}
\label{Section:hardware}

The DUG infrastructure provided us with esentially two tiers of compute capability. $>$2000 nodes were provided based on Intel Xeon Phi `Knights Landing' processors with 64 cores and 128GB RAM, billed at A\$0.35 per node-hour. Also available were eight nodes based on dual socket Intel `Cascade Lake' processors, each with 24 cores and 192 GB RAM, and eight nodes based on dual socket Intel `Ice Lake' processors, each with 16 cores and 1 TB RAM. These higher-performance nodes were billed at A\$2.45 and A\$2.65 per node-hour respectively.

In this paper, these three node types are referred to as `KNL', `CLX' and `ILX'. We use the term 'sockets' to refer to the individual processors within the node. 

With the dual-socket nodes, the default behaviour of the Linux kernel is to spread active processes across both sockets, and it may migrate active processes between sockets to balance load. Although half the node memory is attached to each socket, each has access to the full node memory. However there is a lower latency if sockets are 'pinned' so that the socket uses only its attached memory. This detail is included as background to our later discussion on performance improvement.

The file system at DUG is based on VAST\footnote{https://www.vastdata.com/}, a scalable SSD-based storage accessed via NFS. At the time of this work, DUG's Perth deployment was rated at 90 GB/s read and 15 GB/s write. Compute nodes are connected via NVIDIA Mellanox multi-host NICs, with one card connecting four nodes to the network. The total bandwidth of 50 Gb/s can be rate limiting if multiple nodes access the network concurrently.

\section{Software}
\label{Section:software}
The key software tools for producing image cubes from visibility data were WSClean version 3.0 \citep{2014MNRAS.444..606O} (for imaging) and CASA version 6.1.2.7 \citep{2022PASP..134k4501C} for most data manipulation and arithmetic operations. CASA tasks were scripted in Python, for execution in CASA's IPython application environment. 

WSClean and CASA had been installed by DUG's HPC team prior to the project, for use in an earlier activity. The original implementation of the workflow at the Australian National University (ANU) used MIRIAD \citep{1995ASPC...77..433S} for one task (Feathering). To save time required to resolve difficulties installing a consistent set of dependencies for this package, the task which relied on MIRIAD was allocated to CASA instead, and MIRIAD was not used.

Non-interactive jobs were controlled using RJS, which is DUG's proprietary wrapper for sbatch, which in turn is part of the underlying scheduler SLURM (we used version 23.02.07). RJS provides a simple method for jobs to be executed in parallel, by supplying values for variables in an external schema file.

The operating system kernel was CentOS Linux release 7.7.1908, and RJS jobs were submitted manually via ssh access.

An example set of the RJS scripts and schema files, and ipython scripts, used to implement the workflow are available online \citep{IKImagingGuide}.

During our project, workflow steps were initiated manually, generally with the output of one step being checked before starting the next. Orchestration using a workflow manager was left to be the first stage of full scale deployment in the event that the DUG platform was to be used after our trial was complete.

\section{Workflow}
\label{Section:Workflow}

Key features of the workflow for GASKAP-H\textsc{i} imaging is described in \cite{2022PASA...39....5P}. The workflow had been optimised to run on the AVATAR cluster at the Research School of Astronomy and Astrophysics at the Australian National University (ANU), and a detailed description of the code, along with the logic of each step, is given in the imaging guide \citep{NPImagingGuide}. 

Observations were processed one field at a time. In the current work, initial establishment of the workflow was carried out using SBID 38215, which had previously been processed using ANU facilities. This was done to establish confidence in the modified workflow, by comparison of the resulting images with those previously obtained.

A summary of the workflow steps used at DUG is listed in Table \ref{table:workflow}. As indicated by the rightward column, the process uses some manual steps which were run interactively; some single jobs; and some large jobs which were parallelised, and run concurrently over the largest number of nodes which were available. A graphical representation of the workflow is available as Fig. 2 in \cite{2022PASA...39....5P}.

\begin{table*}[t]
\centering
\begin{tabular}{llll}
\hline
Step & Name & Description & Job Type \\
\hline
01 & Download & Transfer visibility files from CASDA to DUG & Single \\
02 & Bin & Combine (average) frequency channels & Parallel \\
03 & Listobs & Record basic observation metadata & Manual \\
04 & Rotate Ph. Centre & Move all interleaves to common phase centre & Parallel \\
05 & Cont. Sub. & Suppress continuum sources & Parallel \\
06 & Split Channel & Split visibility files 1 per channel & Parallel \\
07 & Make Clean Mask & Create mask where PB $<$ 20\% & Manual \\
08 & Imaging & Create image \& PB for each channel & Parallel\\
10 & Collect images & Consolidate images and PB & Single \\
11 & Import to CASA & Convert from FITS to CASA image format & Single \\
12 & Update Headers & Correct header metadata in CASA images & Single \\
14 & Concatenate & Form image cube \& PB cube & Single \\
15 & Normalise PB & Ensure max value in PB $<$ 1.0 & Single \\
16 & PB correction & Divide image cube by PB cube & Single \\
17 & Get Parkes Cube & Download prior single dish cube & Manual \\
18 & Dropdeg & Ensure ASKAP \& Parkes cubes compatible & Single \\
19 & Update Headers & Ensure ASKAP \& Parkes cubes compatible & Single \\
20 & Smooth & Apply smoothing to the ASKAP cube & Single \\
21 & Feather & Combine ASKAP \& Parkes images & Single \\
\hline
\end{tabular}
\caption{Summary of imaging workflow (More detail is given in \cite{2022PASA...39....5P}). Note: 9 and 13 were not used in the production workflow}
\label{table:workflow}
\end{table*}

Some of the notable features of the workflow, which will be referred to later, are:

\begin{itemize}
\item{`Download': Due to cost of storage it was not feasible to replicate all the visibility data on the commercial platform, nor was this felt a desirable data management practice. Therefore a process step was required to download data from the repository to the working machine immediately prior to processing;}
\item{`Rotate Phase Centre': In this step visibility data from the 108 beams are phase-rotated to a common reference coordinate, which defines the centre of the image. This is a requirement for the joint deconvolution mode in WSClean;}
\item{`Split Channel': To reduce memory demands and processing time during imaging, visibility data are pre-split into individual channels for each of the 108 pointings. Thus, we image each spectral channel separately and in parallel. These images are concatenated into a single cube in a later step;}
\item{`Feather': The physical limitations on spacing of the individual elements of radio interferometers limits the sensitivity to low spatial frequencies. For ASKAP, the shortest 22 m baselines filter out structures that span $>30'$ on the sky, eliminating sensitivity to large angular scales. These missing baselines are commonly referred to as the missing short-spacings. To compensate for these, image cubes from ASKAP are combined with images from the GASS survey obtained with Murriyang, the single-dish Parkes telescope \citep{2009ApJS..181..398M}. The elapsed time for the download of the relevant cube from the Parkes data repository (workflow step 17) was trivial and for this reason we did not attempt to automate it.}
\end{itemize}

\section{Methods and evolution of imaging workflow}
\label{Section:Methods}

Our project contained four distinct stages:
\begin{enumerate}
\item{The working code and workflow were ported from a working installation with minimal changes required to run successfully;}
\item{Experimentation with WSClean parameters to obtain satisfactory images;}
\item{Optimisation of deployment of the code to the compute nodes to improve performance - referred to as `Machine optimisation';}
\item{Optimisation of WSClean parameters to further improve performance - referred to as `software optimisation'.}
\end{enumerate}
These stages are now discussed in more detail.

\subsection{Port of the existing Workflow}
We ported the working code from the ANU implementation, with changes to meet updated science goals and to take best advantage of the DUG platform. Initial trials were conducted using data from SBID 38215.  For the current work we used the full spectral range and resolution, of 2112 channels of 1.157 kHz, and an image size of 4096x4096 7-arcsec pixels. This was a significant increase in computing load compared to previous work, which had used 4-channel binning, giving a total of 528 channels of width 4.628 kHz.

When implementing the workflow at DUG, some technical issues emerged which will be of interest to researchers attempting to replicate our processing on other platforms:
\begin{itemize}
  \item{The first issue was that it was convenient to use DUG's scheduling tool RJS. This improvement required all the job control scripts to be modified to use RJS directives in place of the Portable Batch System commands used at the ANU;}
  \item{In the original workflow, data were binned into 4-channel bins. For improved science output and because of the the additional processing capacity available, we opted to process the data at the full native spectral resolution of 1.157 kHz, so no binning was required. Omission of the binning step led to failure of the phase centre rotation step. This was attributed to the `tiling' scheme in the measurement set; to solve the problem we adopted the simple expedient of reinstating the `bin' operation, but with width 1. This caused the observation data to be re-tiled at a small compute cost and further processing was successful.}
\end{itemize}

\subsection{WSClean parameters}
To provide reference images for the development of the joint deconvolution mode in ASKAPSoft, we produced some cleaned single beam images as well as joint deconvolved (multibeam) images, in order to separate the effects of cleaning and mosaicing. Single beam images were made using beam 21A of SBID 10941, centered on the SMC. These used a single channel (Channel 760 centred on 1419.380 MHz in the LSRK spectral reference frame), with an image size of 1024$\times$1024 7-arcsec pixels. These images were fully contained within the area of the ASKAP footprint (i.e., the combined primary beam map). 

\begin{figure*}[h]
\centering
\scalebox{1.0}{\includegraphics{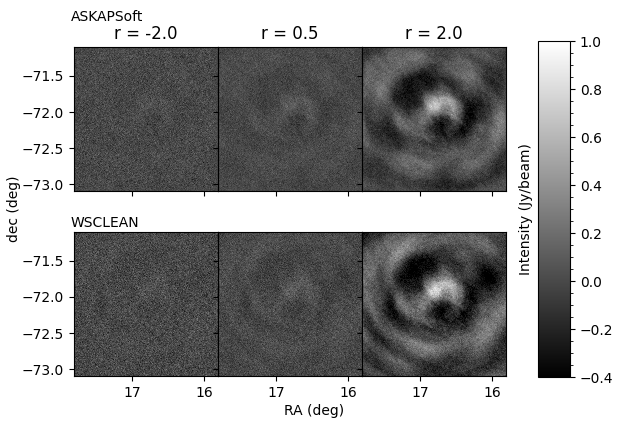}}
\caption{Comparison of dirty images from ASKAPSoft (top row) and WSClean (bottom row) with different Briggs Weighting values. Images are from SBID 10941, single beam 21A, Channel 760, centred at 1419.380 MHz.
\label{fig:10941_dirty_comp}}
\end{figure*}

The key parameters affecting image quality were found to be the clean threshold and the Briggs robustness value, which weights the gridded visibilities to provide a compromise between noise and the final angular resolution. After some experimentation, it was found that optimal images were obtained with a Briggs weighting around 2.0. A comparison for 3 values of the Briggs weighting is given for the single-beam images in Figure \ref{fig:10941_dirty_comp}, showing good qualitative agreement between the two imagers. In both cases, the sidelobe structure becomes more apparent as the weighting transitions from uniform ($-2.0$) to natural ($2.0$), wherein structure from the large amounts of baselines below 2 km begin to dominant the image. Other WSClean parameters used are listed in the first column of Table \ref{table:WSClean}.

For joint deconvolved images of the full field, the image size was initially increased to 5000$\times$5000 7-arcsec pixels (later rounded to 4096$\times$4096). This included an area outside that covered by the joint primary beam map, and noise in this outer zone led to rapidly diverging clean operations. We opted to deal with this by providing a clean mask to exclude parts of the image where the joint beam response was less than 15\% of the peak.

\begin{figure}
\scalebox{0.94}{\includegraphics{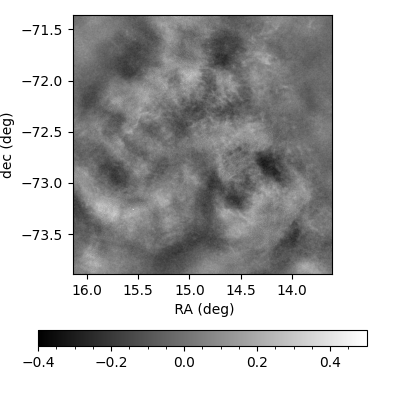}}
\caption{Portion of the cleaned (but not feathered) image cube from SBID 10941 at channel 798. Scale bar is in Jy/Beam.}
\label{fig:fig_10941_798_sub.fits.png}
\end{figure}

Satisfactory image cubes were obtained with a clean threshold of 21 mJy and a Briggs weighting of 1.25. A portion of the cleaned image cube at channel 798 centred on 1419.424 MHz is shown in Figure \ref{fig:fig_10941_798_sub.fits.png}. The image includes the SMC Bar (rightward of RA = 15\textdegree) and Wing (leftward of RA = 15\textdegree), and resolves linear plumes at the upper left, and fine filaments (for example centred on RA = 15.5\textdegree, decl = -72.4\textdegree). For a further discussion of features in the image, please refer to the description of Figure 12 of \cite{2022PASA...39....5P}. WSClean parameters used to produce our Figure \ref{fig:fig_10941_798_sub.fits.png} are listed in the second column of Table \ref{table:WSClean}. Negative intensities in the image are due to the lack of a total power measurement, and are mitigated by feathering with data from the Parkes single dish telescope, at a later stage in the workflow.

For this first successful production imaging, processing time and cost were dominated by the imaging step. Using CLX nodes, imaging consumed 2.9 hr per channel, giving a total of 3,061 node-hours for the full 2112-channel cube (commercial cost $\sim$A\$7,800). Within this, inversion was by far the most expensive process (ie. computation of the image by Fourier Transform of gridded visibility data), consuming typically 2.1 hr per channel. The next most significant cost was the 'Split Channel' step in which separate measurement sets were created for each beam and each channel. Implemented on CLX nodes, this consumed 263 node hours (commercial cost $\sim$A\$670).

\begin{table*}[h]
\centering
\begin{tabular}{lccc}
\hline
Images&Single beam&Initial&Final\\
SBID&10941/21A&10941&38466\\
\hline
-aterm-config&$<$path to beammap.config$>$&$<$path to beammap.config$>$&$<$path to beammap.config$>$\\
-auto-mask&-&-&3\\
-auto-threshold	&-&-&0.3\\
-beam-size&30&30&-\\
-field&0&all&all\\
-fits-mask&-&$<$path to clean mask$>$&$<$path to clean mask$>$\\
-j&$<$Cores per socket / 2$>$&48&$<$Cores per socket / 2$>$\\
-log-time&TRUE&TRUE&TRUE\\
-mask&-&TRUE&TRUE\\
-mgain&0.5&0.7&0.7\\
-multiscale&TRUE&TRUE&TRUE\\
-multiscale-gain&0.1&0.1&0.1\\
-multiscale-scale-bias&0.85&0.85&0.85\\
-multiscale-scales&"0,8,16,32,64,128,256"&"0,8,16,32,64,128,256"&-\\
-name&$<$output name$>$&$<$output name$>$&$<$output name$>$\\
-niter&10,000&10,000&40,000\\
-nmiter&8&5&5\\
-no-negative&TRUE&TRUE&TRUE\\
-no-update-model-required&-&-&TRUE\\
-pol&i&i&i\\
-save-first-residual&TRUE&TRUE&TRUE\\
-scale&7asec&7asec&7asec\\
-size&1024 1024&5000 5000&4096 4096\\
-taper-gaussian&-&-&8\\
-temp-dir&-&-&$<$path to node RAM$>$\\
-threshold&0.010&0.021&0.015\\
-use-idg&TRUE&TRUE&TRUE\\
-weight&briggs [-2, 0.5, 2]&briggs 1.25&briggs 1.0\\
\hline
\end{tabular}
\caption{WSClean parameters, used initially, developed during single beam imaging, and adopted final values. `Final' column relates to timing data in Table \ref{table:38466_final}}
\label{table:WSClean}
\end{table*}

\subsection{Machine optimisation}
Having established a successful run of the workflow, with satisfactory imaging, our next step was to tune the processes to the machine, to reduce processing time and cost. This was done by focusing on three areas, as follows:

\subsubsection{Data transfer}
An issue with using the commercial platform was the need to transfer data from the system of record for processing. To improve the elapsed time and reduce manual intervention, a process was developed based on `drip feed' of the required visibility files for a given observation, with automated restart and throttled to avoid excessive load on CASDA. This process achieved a download time of 6.2 hours for the full data set for a $\sim$10 hr observation (7 TB), without manual intervention.

\subsubsection{Allocation of tasks to software tools}
The software tool used for each step was also reviewed. Opportunities arose in the choice of tools for phase centre rotation (workflow step 4 in Table \ref{table:38466_final}) and for the feathering step (workflow step 21 in Table \ref{table:38466_final}).

For phase centre rotation the chgcentre tool packaged with WSClean was used for the production workflow in preference to CASA. The feathering task was carried out using the CASA `feather' directive, and we were able to eliminate the MIRIAD tool set from the workflow. This was done to simplify the initial setup of the software environment, as mentioned earlier.

\subsubsection{Allocation of tasks to nodes}
An important step was to allocate each workflow task to either KNL nodes or CLX nodes which gave higher performance but were more costly per node-hour. We began this process by allocating the most computationally intensive tasks to CLX. Where short elapsed time could be of benefit (for example where manual intervention is required), tasks were allocated either to CLX or KNL depending on the degree of parallelisation possible. For tasks which were highly parallelisable, KNL was preferred as the machine provided many more KNL nodes than CLX. An example is the 'Split-by-Channel' step which was implemented on KNL nodes despite the higher total cost. Where processing speed was limited by transactions to the file system, tasks were allocated to the lower cost KNL nodes. The final allocation of tasks to node type is shown in Table \ref{table:38466_final}.\\
\\
For two of the most computationally intensive tasks, imaging, and split-by-channel, experiments were carried out to optimise machine performance. We adopted an experimental approach, as it proved difficult to predict the effects of multiple changes to the system configuration on the processing time.

\begin{itemize}
    \item {Concurrent instances of WSClean: Timings and memory high water mark were obtained for runs including multiple concurrent instances of WSClean on the same node (using a single socket). The results in Table \ref{table:10941_node_stack} show an improvement in the run time per image, from 112 to 65 s/image by increasing from one to four concurrent instances. The improvement from a further increase to 8 instances was marginal, so a figure of four instances per socket (ie 8 per dual-processor node) was initially selected.}
    \item{RAM disk: A RAM disk was configured for use by WSClean temporary files, to maintain all processing on the node without accessing the file system.}
    \item{Socket pinning: This technique (outlined in Section \ref{Section:hardware}) was used to increase processing speed, maintaining four instances of WSClean per node, ie. two instances per socket.}
    \item{Split-by-channel: For this operation we opted to use the lower cost KNL nodes and run the split for all beams concurrently, ie 108 processes under the control of 36 jobs. This approach slightly increased the computation cost compared to using CLX nodes, but significantly reduced the overall elapsed time.}
\end{itemize}

\subsection{Software optimisation}
During processing, further opportunities for cost efficiency became apparent, by improved choice of WSClean parameters:
\begin{itemize}
    \item{We were able to reduce file system access by using the `no-update-model-required' parameter, to prevent the updates to the MODEL column within the measurement sets. This achieved a factor of 1.3 improvement in imaging time;}
    \item{File system access was further reduced by specifying the RAM disk location for WSClean's temporary files, using the -temp-dir directive (ensuring a different directory was used for each of the concurrent instances on the same node);}
    \item{Inspection of the WSClean logs showed that larger cleaning scales were not being activated, so the -multiscale-scales parameter was removed. This did not affect imaging time but simplified the imaging script;}
    \item{The multithread parameter to WSCLean (the -j parameter) was set to explicitly match the number of processing threads to the number of cores available on each socket.}
\end{itemize}

These optimisations provided a final imaging time of 39 minutes per channel, an improvement of a factor of 4.5 on the 2.9 hours per channel obtained from the initial port, before optimisation. As shown in Table \ref{table:10941_node_stack}, the RAM disk and socket pinning changes were introduced together, but independent tests indicated that socket pinning made the greater contribution to run time.  Profiling of the final configuration indicated a processor utilisation during imaging of 73\%.

An additional important component of the project is the cost of storage. For our project, storage was charged on a 'high water mark' basis. This was managed by imposing a storage quota, sufficient to cope with all data and intermediate products for one image cube, which ensured that additional costs were not incurred inadvertently. The use of RAM drive during processing, and reductions in the overall processing time each contributed to limiting storage costs.

\begin{table*}
\centering
\begin{tabular}{ccccccccc}
\hline
SBID & Channels&Sockets&Instances&Config&Run time&Run time/image&Max\\
& & & per socket & &(hr)&(min)&memory\\
\hline
10941&0900&1&1&&1.87&112&12GB\\
10941&0904-09071&&4&&4.33&65&\\
10941&0908-0915&1&8&&8.15&61&95GB\\
\hline
38509&All&2&4&No model update&7.62&57&\\
\hline
33486&0915-0918&2&2&RAM disk, socket pinning&3.06&46&\\	
33486&All&2&2&RAM disk, socket pinning&2.62&39&\\
\hline
\end{tabular}
\caption{Imaging times and memory high water mark during experiments with machine and software optimisation. See text for a detailed description}
\label{table:10941_node_stack}
\end{table*}

\begin{table*}[t]
\centering
\begin{tabular}{lllrcccrr}
\hline
Step & Name & Tool & Time per & No. & Elapsed & Node & Cost & Est. Cost \\
& & Used & Job (s) & Jobs & Time (hr) & Type & (node-hr) & (\$A) \\
\hline
01 & Download & Python & 2,443 & 108 & 6.2 & KNL & 73.3 & \\
02 & Bin & CASA & 65,314 & 1 & 18.1 & KNL & 18.1 & \\
03 & Listobs & CASA &  0 & 1 & 0.0 & KNL & 0.0 & \\
04 & Rotate Ph. Centre & WSClean & 515 & 38 & 0.7 & CLX & 2.7 & \\
05 & Cont. Sub. & CASA &211,981 & 2 & 59.9 & CLX & 58.9 & 144\\
06 & Split Channel & CASA & 256,283 & 36 & 38.9 & KNL & 2,562.8 & 897\\
07 & Make Clean Mask & Python & 5,473 & 1 & 1.0 & CLX & 1.5 & \\
08 & Imaging & WSClean & 9419 & 528 & 97.6 & CLX/ILX & 1381.4 & 3522 \\
10 & Collect images & Python & 115 & 1 & 0.0 & KNL & 0.0 & \\
11 & Import to CASA & CASA & 5,203 & 1 & 1.4 & KNL & 1.4 & \\
12 & Update Headers & CASA & 2,137 & 1 & 0.6 & KNL & 0.6 & \\
14 & Concatenate & CASA & 71,104 & 1 & 19.8 & CLX & 19.8 & \\
15 & Normalise PB & Python & 407 & 1 & 0.1 & CLX & 0.1 & \\
16 & PB correction & Python & 732 & 1 & 0.2 & CLX & 0.2 & \\
17 & Get Parkes Cube & Manual & 0 & 0 & 0.0 & - & 0.0 & \\
18 & Dropdeg & CASA & 263 & 1 & 0.1 & CLX & 0.1 & \\
19 & Update Headers & CASA & 7 & 1 & 0.0 & CLX & 0.0 & \\
20 & Smooth & CASA & 35,195 & 1 & 9.8 & CLX & 9.8 & \\
21 & Feather & CASA & 43,185 & 1 & 12.0 & CLX & 12.0 & \\
\hline
\multirow{2}{35pt}{TOTALS} & & & & & 63.2 & KNL & 2,656.3 & 930\\
& & & & & 199.7 & CLX/ILX & 1486.4 & 3,790\\
\hline
\multicolumn{3}{l}{GRAND TOTAL (Excluding storage costs)} & & & 262.9 &  &  & 4,720\\
\hline
\end{tabular}
\caption{Processing cost for image cube production after optimisation, SBID 38466.}
\label{table:38466_final}
\end{table*}

\begin{figure*}[h]
\scalebox{0.6}{\includegraphics{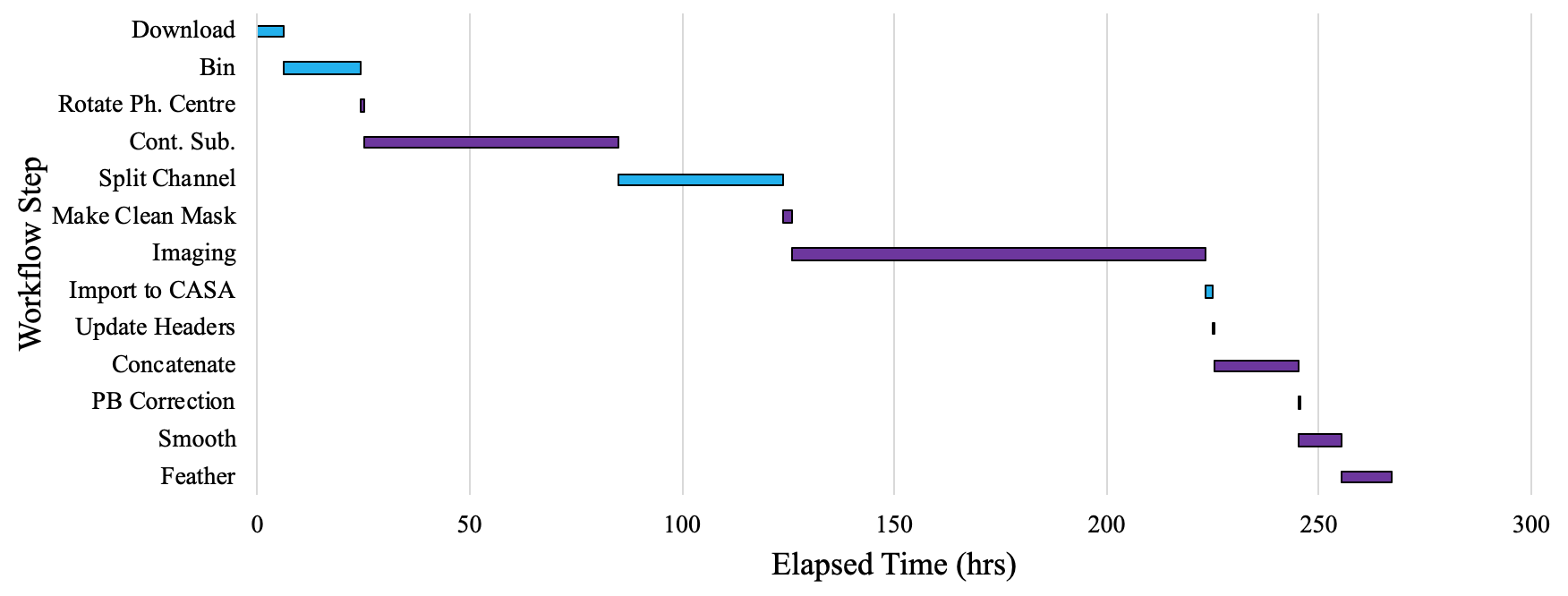}}
\caption{Gantt chart for non-trivial workflow steps, SBID 38466. Light blue bars are KNL processing, dark purple bars are CLX processing.
\label{fig:38466_gantt}}
\end{figure*}

\section{Results} \label{Section:Results}
After successfully porting the workflow to the commercial provider's infrastructure, machine optimisation and software parameter optimisation significantly reduced the cost per image cube, by reducing both computational and storage components of the cost.

Compute times and elapsed time for the full workflow after optimisation are listed in Table \ref{table:38466_final}, for the processing of SBID 38466. The cost in node-hr is the product of the time per job and the number of jobs, divided by two if the job is run on a dual-socket processor without socket pinning.  The total estimated cost per image cube is A\$4,720, and the table lists costs for individual process steps valued at more than A\$100.

The Gantt chart in Figure \ref{fig:38466_gantt} illustrates the major contributors to the overall elapsed time of 11 days, 3 hours, per image. The peak storage of around 25 TB per cube had a commercial cost of $\sim$A\$1,600 per month, which translates to a production imaging cost of A\$530 per 10 hr integration field, provided that multiple workflows can be managed at 3 per month.  Thus the total processing cost for GASKAP-H\textsc{i} full survey images (including processing and storages costs) is estimated as A\$5,250 per field.

Having met the technical goals of the project, production imaging was carried out until the project budget was fully consumed. The optimised workflow was employed to image three further fields from the Pilot II survey, (SBIDs 38215, 38509 and 38466). Thus four image cubes were supplied to the GASKAP-H\textsc{i} science team. 

The modified workflow, developed in this study, is documented in an updated imaging guide, available online with the code \citep{IKImagingGuide}. The guide referred to here is to be used in conjunction with the more comprehensive guide by \cite{NPImagingGuide}.

The images produced have contributed to the development of joint deconvolution in ASKAPSoft, which is expected to ultimately become the primary system for producing GASKAP images.

In the next section we document `lessons learned', which may help inform other teams wishing to use commercial supercomputing for large scale processing. 

\section{Lessons Learned} \label{Section:Lessons}
Incidents and their resolution were documented throughout the project. At completion of the project, authors who had been key participants from the data processing team and the HPC provider, met to review lessons learned and prioritise them. The lessons are summarised in Table \ref{table:lessons_learned}, ranked by importance as determined by a majority voting.

\begin{table*}
\centering
\begin{tabular}{ccl}
\hline
Item & Priority & Lesson \\
\hline
1 & H & Budget (financially \& in schedule) for contingencies \& re-runs \\
2 & H & Develop QA script/process to evaluate runs \\
3 & H & Know your budget \& provider's cost structure to maintain plan within your budget \\
4 & H & Regular team meetings \& scheduling of work more valuable than on-demand support \\
5 & H & Ensure processes are documented \& software tools and special processes are preserved for future projects \\
6 & H & Beware of changes of aims / scope - e.g. from software development to production processing \\
\hline
7 & M & Understand the storage costs - make provision for `offline' storage if possible \\
8 & M & Don't use commercial time for development \& testing: Have core algorithms already developed and tested \\
9 & M & (provider) Understand client requirements before committing time/resources (eg optimising scripts) \\
10 & M & Build parameter testing into the process (inc. explore software versions) \\
\hline
11 & L & Decide in advance what tools to use, and have a standard environment specified or ready to install  \\
12 & L & If on a fixed budget, ensure that budget spend is transparent and visible \\
13 & L & Need resource costs/charging to be aware of `spiky' workload (eg peak based accounting) \\
14 & L & Also spiky availability of data and/or people to QA the data (makes peak based accounting worse) \\
15 & L & Need communication and timely follow-up to supplement regular meetings \\
17 & L & Be open to suggestions on how to optimize algorithms and code \\
19 & L & Have an advance strategy for storage / management of data \\
20 & L & Have clear project structure including who is the customer \\
\hline
\end{tabular}
\caption{Summary of lessons learned during the project}
\label{table:lessons_learned}
\end{table*}

Key lessons are expanded as follows:
Lessons 1, 3, 6, 8, 9 and 10 relate to the need to plan the activity in advance, in particular to consider to what extent the port to a commercial provider will include code optimisation or development. Our experience is that some code modification is almost inevitable, and optimisation work is highly desirable. However, extended development work can incur fixed costs including access fees or storage costs, which can eat into the project budget;

Lesson 2 is a suggestion to reduce elapsed time and compute time spent in development / optimisation. As an example, in the current project we used single-beam images to obtain imaging parameters at considerably lower cost than in processing full cubes;

Lessons 4 and 15 arise from our finding that it was extremely helpful for progress and issue resolution to have regular catchups between the astronomers and the provider's HPC team. We had a mechanism for ad-hoc support but the team meetings were valuable in being able to talk issues through in more detail and foreshadow upcoming work;

Lessons 5 and 11 are related to future deployment of the optimised workflow. It is very helpful if the environment and tool set is defined so that it can be installed rapidly at a new provider;

Lessons 7, 13, 14, and 19 relate to the terms of the contract with the commercial provider. The cost structure may include fixed monthly charges, charges for storage or high water mark storage, in addition to compute time. Also the service catalogue may provide low cost 'offline' storage. It's important for the parties to agree on a set of services and a cost structure with works for the astronomers and the commercial partner;

Lessons 12 and 20 relate to ongoing project management, and the importance of ensuring that progress and consumption of the budget are coordinated, so that project decisions can be made in a timely manner; and

Lesson 17 reinforces the benefits which can be obtained by the astronomers and the HPC team working together with a free exchange of ideas.

\section{Discussion}
 During the project we identified that the use of commercial supercomputing had some clear advantages and disadvantages. The main advantage is that the processing infrastructure was available on demand; there was no merit-based application process and no wait for access to the machine. Due to high availability, we encountered no delays in running jobs. The main disadvantage was the need to port the workflow to the new platform, as detailed in this paper, although this was mitigated by extensive assistance from the provider's HPC team.  Other important considerations are data management and cost. On data management, the primary issue is the need to feed data from the repository of record, and the costs associated with large scale storage at the commercial provider. More generally in terms of costs, it is necessary for the project to have available budget to fund work at the commercial provider.

 An important step was to allocate part of the project budget to experiments to improve the processing time and cost. Our improvements were a mix of changes to imaging parameters, and changes to the deployment of the software on the machine. The latter optimisations undoubtedly require good knowledge of the supercomputing system, and in our case we were able to obtain this expertise from the provider's own HPC team. As a generic lesson, we recommend researchers take action to have this type of expertise available, either from the provider or attached to their own research team.

Documentation of our optimised workflow enabled us to provide a resource estimate for processing GASKAP-H\textsc{i} full survey data at DUG: a cost of $\sim$A\$5,250 and an elapsed time of 11 days per field.

There is doubtless further scope for optimisation of the workflow, using techniques in both machine and software optimisation. For future work it is a matter for the project team to decide how much time and compute resource to devote to the additional optimisation process itself. 

Researchers should also be aware of enhancements to the infrastructure which may become available during the project. For example during the project discussed here, the commercial provider had already worked to upgrade the machine with the addition of new nodes with improved memory and performance, so permitting further improvements in cost and time.

An additional point is that our optimisation approach can usefully be applied to migrations to non-commercial platforms; as shown, significant improvements in efficiency may be possible, which can reduce turnaround time and improve the productivity when processing on a publicly-owned facility, even though users may not see defined dollar costs associated with these resources.

\section{Conclusion}
In the project our key aim of trialling commercial supercomputing was met, along with the four sub-objectives laid out in the introduction:

\begin{enumerate}
    \item{We have demonstrated the feasibility of using commercial infrastructure for processing data from the GASKAP-H\textsc{i} pilot and full surveys;}
    \item{We have obtained estimates of cost and elapsed time for producing image cubes from the future full survey;}
    \item{We used the joint deconvolution technique with WSClean to produce reference images to assist development of the new algorithm in ASKAPSoft; and}
    \item{contributed four image cubes for use by the science team.}
\end{enumerate}

Our documentation of the process, and lessons learned, will assist researchers in taking advantage of commercial processing for GASKAP and perhaps other surveys.

\section{Acknowledgements}
The authors thank CSIRO, DUG Technology, and Curtin University for supporting this work financially and in kind.

This scientific work uses data obtained from Inyarrimanha Ilgari Bundara / the Murchison Radio-astronomy Observatory. We acknowledge the Wajarri Yamaji People as the Traditional Owners and native title holders of the Observatory site. CSIRO’s ASKAP radio telescope is part of the Australia Telescope National Facility (https://ror.org/05qajvd42). Operation of ASKAP is funded by the Australian Government with support from the National Collaborative Research Infrastructure Strategy. ASKAP uses the resources of the Pawsey Supercomputing Research Centre. Establishment of ASKAP, Inyarrimanha Ilgari Bundara, the CSIRO Murchison Radio-astronomy Observatory and the Pawsey Supercomputing Research Centre are initiatives of the Australian Government, with support from the Government of Western Australia and the Science and Industry Endowment Fund.

This paper includes archived data obtained through the CSIRO ASKAP Science Data Archive, CASDA (http://data.csiro.au)

\section{ORCID IDs}
\noindent
Ian Kemp https://orcid.org/0000-0002-6637-9987 \\
Nickolas Pingel https://orcid.org/0000-0001-9504-7386 \\
Steven Tingay https://orcid.org/0000-0002-8195-7562 \\
Daniel Mitchell https://orcid.org/0000-0002-1828-1969 \\
James Dempsey https://orcid.org/0000-0002-4899-4169 \\
Helga Dénes https://orcid.org/0000-0002-9214-8613 \\
John Dickey https://orcid.org/0000-0002-6300-7459 \\
Steven Gibson https://orcid.org/0000-0002-1495-760X \\
Kate Jameson https://orcid.org/0000-0001-7105-0994 \\
Callum Lynn https://orcid.org/0000-0001-6846-5347 \\
Yik Ki Ma https://orcid.org/0000-0003-0742-2006 \\
Antoine Marchal https://orcid.org/0000-0002-5501-232X \\
Naomi McClure-Griffiths https://orcid.org/0000-0003-2730-957X \\
Snežana Stanimirović https://orcid.org/0000-0002-3418-7817 \\
Jacco van Loon https://orcid.org/0000-0002-1272-3017 \\

\bibliographystyle{elsarticle-harv} 
\bibliography{KempEtAl.bib}

\end{document}